\documentclass[floatfix,prb,aps,twocolumn,showpacs,preprintnumbers,amsmath,amssymb]{revtex4}  
\usepackage{graphicx} 
\usepackage{dcolumn} 
\usepackage{exscale} 
\usepackage{bbm}  
\usepackage{amsfonts} 
 
\def\shiftdown#1{#1\llap{\lower.04ex\hbox{#1}}}

\newcommand{\I}{ {\rm Im}} 
\newcommand{\R}{ {\rm Re}} 
\begin{document} 
\title{Theory of strong localization effects of light in disordered  
loss or gain media} 
 
\author{Regine Frank$^{1,2}$,
        Andreas Lubatsch$^{1,3}$, and 
        Johann Kroha$^1$} 
\affiliation{$^1$Physikalisches Institut, Universit\"at Bonn, 
                 53115 Bonn, Germany\\
             $^2$Institut f\"ur Astronomie und 
                 Astrophysik,  
                 Universit\"at T\"ubingen, 72076 T\"ubingen, Germany\\  
             $^3$Laboratoire de Physique et Mod\'elisation des 
                 Milieux Condens\'es/CNRS, 
                 38042 Grenoble, France} 
\begin{abstract} 
\noindent 
We present a systematical theory for the interplay of strong   
localization effects and absorption or  
gain of classical waves in 3-dimensional, disordered dielectrics.  
The theory is based on the selfconsistent Cooperon resummation,  
implementing the effects of energy conservation  
and its absorptive or emissive corrections by an exact,  
generalized Ward identity. 
Substantial renormalizations are found, depending on whether the  
absorption/gain occurs in the scatterers or in the background medium. 
We find a finite, gain-induced correlation volume which may be significantly  
smaller than the scale set by the scattering mean free path, 
even if there are no truly localized modes. Possible consequences for 
coherent feedback in random lasers as well as the 
possibility of oscillatory in time behavior induced by sufficiently 
strong gain are discussed. 
\end{abstract} 
\maketitle 

\section{Introduction}

The strong or Anderson localization (AL) \cite{And58} of light in 
$d=3$ dimensional, random media has remained a fascinating and controversial 
issue, despite its early theoretical anticipation \cite{John83,Kro93} and  
despite the experimental observation of its precursor effect, the  
coherent backscattering light cone of weak localization.
\cite{Kug84} Early experimental reports of an   
anomalously low diffusion constant in strongly scattering powders \cite{Alb91} 
were explained by a reduction of the transport velocity by  
Mie resonances. \cite{Tig92}  
However, recent works on the experimental proof of  
light localization \cite{Wie97} has lead to a 
controversial, still ongoing debate. \cite{Sch99,Wie99}  
The reason resides in the difficulty to distinguish AL 
from absorption in the medium. The problem becomes even more pressing for 
localization in random media with stimulated emission  
({\em random lasers}). \cite{Cao03b} Here the scattering mean free  
path is large compared to the wave length, $\ell \gg \lambda$, so that 
localization effects are expected to be small. However, the experiments  
indicate unambiguously, by direct observation \cite{Cao00} and especially  
by measuring the photon statistics, \cite{Cao01STAT} that the laser emission
is due to coherent feedback and occurs from spatially confined spots  
in the sample. As an alternative to lasing from Anderson localized modes, 
the existence of preformed cavities in the random medium has been proposed 
in Ref.\ \onlinecite{Apalkov02}. 

AL has been understood 
\cite{gangof4,Vol80} as an effect of repeated self-interference  
(so-called Cooperon contributions) of diffusive modes. Since diffusion, as 
a hydrodynamic phenomenon, relies on particle number or energy conservation,
this raises the fundamental question for the 
fate of AL in active media altogether. 
Intuitively one expects coherent amplification to enhance the transmission,
but at the same time also the Cooperon interference. Since the latter tends to 
localize a wave and hence to reduce the transmission, the overall 
effect is left an open issue. Paasschens {\it et al.} \cite{Beenakker96},
treating the transmission in a linearized way, prove analytically in $d=1$,  
that amplification diminishes the transmission just like absorption does. 
In contrast, 
Jiang {et al.} \cite{Soukoulis99}, going beyond the linear approximation,  
find numerically an enhanced transmission in the long-time limit.
Although important insight has been obtained by further numerical
studies in $d=1$ \cite{Cao04} and $d=2$, \cite{Cao05} 
this controversial situation calls for a semi-analytical theory to
systematically analyze transport quantities like 
the diffusion coefficient $D$ and the intensity correlation length 
$\xi_a$. This is especially so
in $d=3$ where the realization of AL modes is unclear
even in the passive system.

In this article we present such a theory, based on the selfconsistent 
Cooperon resummation pioneered by Vollhardt and W\"olfle.  
\cite{Vol80} It allows to distinguish AL properties from
absorption/gain-induced decay or growth. 
The evaluations will be done for $d=3$, although the theory is valid
for $d=1,2,3$.
As the central results we 
(1) recover, for small gain, the duality of AL with respect to 
absorption or gain, together with a spatially homogeneous, 
exponential decay/growth in the long-time limit, 
thus reconciling the results of Refs.\ \onlinecite{Beenakker96,Soukoulis99}.
(2) In addition, we find strong loss or gain dependent renormalizations 
of $D$ and of $\xi_a$, asymmetrical in loss or gain, 
originating not from wave interference but from the violation of the 
conservation laws. These renormalizations are, in particular, sensitive to 
whether the loss (gain) occurs in the homogeneous background medium or 
in the random scatterers, and thus do not occur in $d=1$ layered structures.
(3) We predict that due to loss or gain 
the intensity correlation length can be
substantially smaller than the scattering 
mean free path $\ell$, constituting a finite coherence volume.   
Possible implications on coherent feedback
in random lasers are discussed.

\section{Model and Transport Theory}

We consider a system of randomly positioned spherical scatterers   
in a background medium with dielectric constants  
$\epsilon_s$ and $\epsilon_b$, respectively. On the semiclassical level, 
linear absorption and the stimulated part of emission are represented by 
a positive or negative imaginary part of the index of refraction,  
$n_s=\sqrt{\epsilon_s}$, $n_b=\sqrt{\epsilon_b}\, $. 
Since we will focus on the fundamental problems 
posed above rather than a quantitative description, we will neglect the 
vector nature (polarization) of light and consider scalar, 
classical waves only. These obey the wave equation 
\begin{equation} 
\label{eq:field} 
\frac{\omega^2}{c^2} \, \epsilon (\vec{r}\,) \Psi_\omega(\vec{r}\,) 
+ \nabla ^2 \Psi_\omega( \vec{r}\,) 
= -i \omega \frac{4\pi}{c^2}  j_\omega(\vec{r}\,)\ , 
\end{equation} 
where $c$ denotes the vacuum speed of light  
and  $j_\omega (\vec{r}\,)$ an external source. 
The dielectric constant  
$ \epsilon(\vec{r}) = \epsilon_b + \Delta\epsilon\, V(\vec{r}\,)$,  
$\Delta\epsilon = \epsilon_s - \epsilon_b$, 
describes the arrangement of scatterers through the function 
$V(\vec{r}\,) = \sum_{\vec{R}} S_{\vec{R}}\,(\vec{r}-\vec{R}\,)$, with 
$S_{\vec{R}}\,(\vec{r}\,)$ a localized shape function 
at random locations $\vec{R}$. 
The Fourier transform  
of the retarded, disorder averaged Green's function of Eq.\ (\ref{eq:field})
reads,
\begin{equation}
\label{Gk}
G_{\vec{k}}^{\omega} = \frac{1}
{\epsilon_b (\omega /c)^2 - \vert \vec{k} \vert^2 - \Sigma^{\omega}_{\vec{k}}} \ ,
\end{equation} 
where the retarded selfenergy $\Sigma _{\vec{k}} ^{\omega}$  
arises from scattering off the random ``potential'' 
$-(\omega /c)^2(\epsilon_s - \epsilon_b )V(\vec{r}\,)$.
The mode density is given by 
$N(\omega)=-(\omega/\pi)\I G_0^{\omega}$,  
$G_0^{\omega}\equiv \int d^3k/(2\pi)^3\, G_{\vec{k}}^{\omega}$.

To develop a transport theory
we now turn to the 4-point intensity correlation function, 
defined in terms of the non-averaged Green's functions 
$\hat G$, $\hat G^*$ and the disorder average $\langle \dots \rangle$ as 
$\Phi^{\omega}_{\vec{q}\vec{q}^{\prime}}(\vec{Q},\Omega)= 
\langle \hat G^{\omega_+}_{{\vec{q}}_+{\vec{q}}_+^{\,\prime}} 
        \hat G^{\omega_-\, *}_{{\vec{q}}_-^{\,\prime} {\vec{q}}_-}  
\rangle$.
It is determined by the kinetic equation, see, e.g., 
Ref.\ \onlinecite{Lubatsch05}, 
\begin{eqnarray} 
\label{boltzmann} 
\left[ 
\omega\Omega\frac{\R{\epsilon_b}}{c^2} 
- Q\, (\vec{q}\cdot\hat{Q}) +\frac{i}{c^2\tau ^2} 
\right] 
\Phi^{\omega}_{\vec{q}\vec{q}^{\prime}} 
&=& \nonumber\\ 
&&\hspace*{-4.3cm}- i \I  G^{\omega}_{\vec{q}} 
\left[ 1 + 
\left\lmoustache  \frac{{\rm d}^3 q^{\prime\prime}}{(2 \pi)^3} \right. 
 \gamma^{\omega}_{\vec{q}{\vec{q}^{\prime\prime}}} 
\Phi^{\omega}_{\vec{q}^{\prime\prime}\vec{q}^{\prime}} 
\right]. 
\end{eqnarray} 
Here we have introduced the usual \cite{Lubatsch05}  
center-of-mass and relative frequencies and momenta: 
The variables $\Omega$, $\vec Q$ are associated with the time and 
position dependence of the averaged energy density, with
$\hat Q =\vec Q/|\vec Q|$, while  
$\omega_{\pm} = \omega \pm \Omega /2$ and  
$\vec{q}_{\pm} = \vec{q} \pm \vec{Q}/2$ etc. are the frequencies  
and momenta of in- and out-going waves, respectively. 
In order to analyze the correlation function's long-time 
($\Omega \to 0$) and long-distance ($ \vert \vec{Q} \vert \to 0$)  
behavior, terms of $O(\Omega^2, Q^3, \Omega Q)$ have been  
neglected here and throughout this paper.  
Eq.\ (\ref{boltzmann}) contains both, the {\em total}  
quadratic momentum relaxation rate $1/\tau^{2}=c^2\,\I ( \epsilon_b  
\omega^2/c^2-\Sigma ^{\omega})$ (due to absorption/gain in the  
background medium as well as impurity scattering) and the 
irreducible two-particle vertex function 
$\gamma^{\omega}_{\vec{q} \vec{q}^{\,\prime}}(\vec{Q},\Omega)$.
The energy conservation is implemented in a field theoretical sense  
by a Ward identity (WI) which has been derived for the photonic case in 
Ref.\ \onlinecite{Lubatsch05}, and which for scalar waves takes the exact 
form 
\begin{eqnarray} 
\label{Ward_2} 
\Sigma^{\omega_+}_{\vec{q}_+} - \Sigma^{\omega_-\, *}_{\vec{q}_-} 
 \!\!&-& \!\! 
\left\lmoustache  \!\!\frac {{\rm d}^3 q^{\prime}}{(2\pi)^3} \right. 
\left[G^{\omega_+}_{{\vec{q}}_+^{\,\prime}} - G^{\omega_-\, 
  *}_{{\vec{q}}_-^{\,\prime}} \right] \,  
{ \gamma}^{\omega}_{{\vec{q}}^{\,\prime}{{\vec{q}}}}({\vec{Q}},\Omega) 
\\ 
 \! \!\!&=& \!\!\! 
f_{\omega}(\Omega)\! 
\left[\! 
\R{\Sigma}^{\omega}_{{\vec{q}}} 
 \!+\!\! 
\left\lmoustache  \!\!\frac {\rm d^3 q^{\prime}}{(2\pi)^3} \right. 
\R {G}^{\omega}_{{\vec{q}}^{\,\prime}} \,  
{ \gamma}^{\omega}_{{\vec{q}}^{\,\prime}{{\vec{q}}}}({\vec{Q}},\Omega) 
\right]\! . 
\nonumber 
\end{eqnarray} 
The right-hand side of Eq.\ (\ref{Ward_2}) represents reactive effects 
(real parts), originating from the explicit $\omega^2$-dependence of the 
photonic random ``potential''. In conserving 
media ($\I \epsilon _b = \I \epsilon _s  =0$) these terms renormalize the 
energy transport velocity $v_{\mbox{\tiny E}}$ relative to the 
average phase velocity  $c_p$  without 
destroying the diffusive long-time behavior.\cite{Kro93,Lubatsch05} 
In presence of loss 
or gain, however, these effects are enhanced via the prefactor  
$f_{\omega}(\Omega)= 
(\omega\Omega \R \Delta\epsilon + i\omega^2 \I \Delta\epsilon )/ 
(\omega^2     \R \Delta\epsilon + i\omega\Omega \I \Delta\epsilon )$, 
which now does not vanish in the limit 
$\Omega \to 0$. As a consequence, more severe renormalizations of the 
diffusion coefficient $D$ and a finite intensity correlation  
length $\xi _a$ are induced, as seen in section \ref{selfconsistent}.

\section{Selfconsistent Solution}
\label{selfconsistent}
As long as wave interference effects are not taken into account, the
selfenergy and the irreducible vertex may be evaluated within a
local approximation, like the independent scatterer approximation 
employed in Ref.\ \onlinecite{Tig93} or modifications of 
the coherent potential approximation (CPA). \cite{Bus95} 
In this case, these quantities become $q$-independent,
$\Sigma _{\vec{q}} ^{\omega} \to \Sigma _{0} ^{\omega}$,
${\gamma}^{\omega}_{{\vec{q}}^{\,\prime}{{\vec{q}}}}({\vec{Q}},\Omega)
\to \gamma _0^{\omega}$, and Eq. (\ref{boltzmann}) may be solved 
in a straight-forward way \cite{Lubatsch05} with
the help of Eq. (\ref{Ward_2}). However, to account for Anderson
localization effects, the $q$-dependence 
of ${\gamma}^{\omega}_{{\vec{q}}^{\,\prime}{{\vec{q}}}}({\vec{Q}},\Omega)$
is known to be essential. \cite{gangof4,Vol80}
To solve the kinetic equation (\ref{boltzmann}) for this case, 
$\Phi^{\omega}_{\vec{q}\vec{q}^{\prime}}$ 
is expanded \cite{Kop84} into its moments with respect to the longitudinal 
current vertex $(c_p \vec p \cdot \hat Q)$, i.e. into 
the energy density correlator 
$P^{\omega}_{\mbox{\tiny E}}(\vec{Q}, \Omega)$ 
and the energy current density  
correlator $J^{\omega}_{\mbox{\tiny E}}(\vec{Q}, \Omega)$, 
\cite{Kro93,Lubatsch05}
\begin{eqnarray}
P^{\omega}_{\mbox{\tiny E}}(\vec{Q}, \Omega)
&=&\left(\frac{\omega}{c_p}\right)^2 \int \frac{d^3q\, d^3q'}{(2\pi)^6} \
\Phi^{\omega}_{\vec{q}\vec{q}^{\prime}}(\vec{Q},\Omega)\\
J^{\omega}_{\mbox{\tiny E}}(\vec{Q}, \Omega)
&=&\frac{\omega v_{\mbox{\tiny E}}}{c_p} \int \frac{dq\, dq'}{(2\pi)^6} \
(\vec p \cdot \hat Q)\ \Phi^{\omega}_{\vec{q}\vec{q}^{\prime}}(\vec{Q},\Omega) \ .
\end{eqnarray}
In these definitions the phase velocity $c_p$ 
of the disordered medium is defined via the
dispersion obtained from the averaged single-particle Green's function,
Eq.\ (\ref{Gk}), as
\begin{equation}
c_p = \R \frac{c}{\sqrt{\epsilon_b-\Sigma _0^{\omega}c^2/\omega^2)}} \ ,
\end{equation}
and the transport velocity $v_{\mbox{\tiny E}}$ 
will be given below.
Inserting the expansion by moments into the kinetic equation 
(\ref{boltzmann}) and
employing the WI, one obtains the exact continuity equation and the current  
relaxation equation, respectively, 
\begin{eqnarray} 
\Omega P^{\omega}_{\mbox{\tiny E}} + Q J^{\omega}_{\mbox{\tiny E}} =  
\frac{4\pi i \,\omega\, N(\omega )} 
       {g^{(1)}_{\omega}\left[ 1 + \Delta(\omega) \right] c_p^{2}} 
\!\!&+&\!\! \frac{i [g^{(0)}_{\omega} + \Lambda(\omega) ]} 
       {g^{(1)}_{\omega}\left[ 1 + \Delta(\omega) \right] } 
       P^{\omega}_{\mbox{\tiny E}}  
\nonumber\\  
\label{continuityL}\\ 
\label{CDR} 
\left[\omega\Omega\frac{\R{\epsilon_b}}{c^2} 
      +\frac{i}{c^2\tau ^2}+iM(\Omega) 
\right] 
J^{\omega}_{\mbox{\tiny E}} \!\!&+&\!\! 
\tilde A\, Q P_{\mbox{\tiny E}}^{\omega} =0\ , 
\end{eqnarray} 
The expression for the transport velocity is now determined by the 
condition that the density and the current correlators must 
appear on the left-hand side of the continuity equation (\ref{continuityL})
without additonal prefactors. One obtains, \cite{Kro93}
\begin{equation}
v_{\mbox{\tiny E}} = \frac{c\ (c/c_p)}
{1-\R(\gamma_0^{\omega} G_0^{\omega} + \Sigma_0^{\omega})} \ .
\end{equation}
The expressions for the coefficients $\tilde A$, $\Delta$, $\Lambda$, 
$g_{\omega}^{(0)}$, $g_{\omega}^{(1)}$ follow in analogy 
to Ref.\ \onlinecite{Lubatsch05},
\begin{eqnarray}
\tilde A
&=& \frac {c_p v_{\mbox{\tiny E}}} {\omega}
\left[
1-\frac { \omega^2 \I \epsilon_b} {u_{\epsilon} \I G_0}
+
\frac{r_{\epsilon}A_{\epsilon}}
{u_{\epsilon} \I G_0}
\right]
\nonumber\\
\Delta(\omega)
&=& B_{\epsilon}A_{\epsilon} +i
r_{\epsilon}\partial_{\Omega}A_{\epsilon}(\Omega) \nonumber \\
\Lambda (\omega)
&=&
i \omega^2 \I \epsilon_b
-i r_{\epsilon} A_{\epsilon}
\nonumber\\
g_{\omega}^{(0)}
&=&
\frac {2\omega}{c^2}\I \epsilon_b
\nonumber\\
g_{\omega}^{(1)}
&=&
\frac {4\omega}{c^2}\R \epsilon_b
\nonumber
\end{eqnarray}
They enter  
into the transport quantities given explicitly below.  
Eq.\ (\ref{continuityL}) represents the local energy conservation 
with corrections due to absorption or gain as well as reactive  
corrections. Eq.\ (\ref{CDR}) describes the time evolution of the 
current density ($J^{\omega}_{\mbox{\tiny E}}$) induced by a  
density distribution ($P_{\mbox{\tiny E}}^{\omega}$) due to 
momentum relaxation and the memory kernel $M(\Omega)$.  
$M(\Omega)$ is a functional of the irreducible 2-wave vertex  
$\gamma^{\omega}_{{\vec{q}}^{\,\prime}{{\vec{q}}}}(Q,\Omega)$.  
Its characteristic $\Omega$-dependence describes  
memory effects due to the diffusion and interference as well as 
amplification or absorption of waves in the medium, see below.  
Eqs.\ (\ref{continuityL}), (\ref{CDR}) are easily solved to obtain 
the diffusion pole form for the density correlator, 
\begin{eqnarray} 
\label{P_E} 
P_{\mbox{\tiny E}}^{\omega}(Q,\Omega) = 
\frac{4\pi i N(\omega) /  
(g^{(1)}_{\omega}\left[ 1 + \Delta(\omega) \right] c_p^{2})} 
{\Omega + i Q^2 D + i \xi_a^{-2}D}\ , 
\end{eqnarray} 
with the $\Omega$-dependent diffusion coefficient $D(\Omega)$, 
\begin{eqnarray} 
\label{D_omega_full} 
D(\Omega)  
\left[1 - i \, \Omega \omega \tau ^2 \R\epsilon_b  \right] =   
D_0^{tot} -c^2\tau^2 D(\Omega ) M(\omega ) 
\end{eqnarray} 
and the absorption or gain induced correlation  
length $\xi_a$ of the diffusive modes, 
\begin{eqnarray} 
\label{xi_a} 
\xi_a^{-2}&=& 
\frac 
{r_{\epsilon} A_{\epsilon} -2\omega^2\I \epsilon_b} 
{2\R\epsilon_b-A_{\epsilon}B_{\epsilon}/\omega}\ 
\frac {1} {\omega D(\Omega)}. 
\end{eqnarray} 
The diffusion constant without memory effects,  
$D_0^{tot}=D_0+D_b+D_s$, consists of the bare diffusion  
constant, \cite{Kro93} 
\begin{eqnarray} 
D_0 = \frac  
{2v_{\mbox{\tiny E}} c_p} 
{ \pi N(\omega)} 
\left\lmoustache  \!\!\frac {{\rm d}^3 k}{(2\pi)^3} \right. 
[\vec{k}\cdot\hat{Q}]^2 (\I G_{\vec{k}}^{\omega})^2 
\label{Dbare} 
\end{eqnarray} 
and renormalizations from absorption or gain in the background medium 
($D_b$) and in the scatterers ($D_s$), 
\begin{eqnarray} 
\label{D_B} 
D_b &=&  \frac 14\left(\omega\tau \right)^2  \, \I\epsilon_b\, \tilde{D}_0 \\
\label{D_S}
D_s &=&  \frac 18 r_{\epsilon}A_{\epsilon}\tau^2 \tilde{D}_0, 
\end{eqnarray} 
where $\tilde D_0$ is the same as in Eq.\ (\ref{Dbare}), 
with $(\I G_{\vec{k}}^{\omega})^2$ replaced by 
$(\R G_{\vec{k}}^{\omega\,2})$. In~Eqs.~(\ref{xi_a})-(\ref{D_S}) 
the following short-hand notations have 
been introduced, 
\begin{eqnarray}
u _{\epsilon}&=&\frac {\I (\Delta\epsilon \Sigma^{\omega})}
                 {\I (\Delta\epsilon G_0^{\omega}) }
\ ,\qquad\qquad
r_{\epsilon} = {\I \Delta\epsilon}/{\R \Delta\epsilon},
\nonumber\\
A_{\epsilon} &=& 2 [u _{\epsilon} \R G_o +  \R \Sigma_o] \nonumber\\
B_{\epsilon} &=& \frac{(\R\Delta\epsilon)^2+(\I\Delta\epsilon)^2}
{2\omega^2(\R\Delta\epsilon)^2}.
\nonumber
\end{eqnarray}

\begin{figure}[t] 
\begin{center} 
\includegraphics[width=0.9\linewidth]{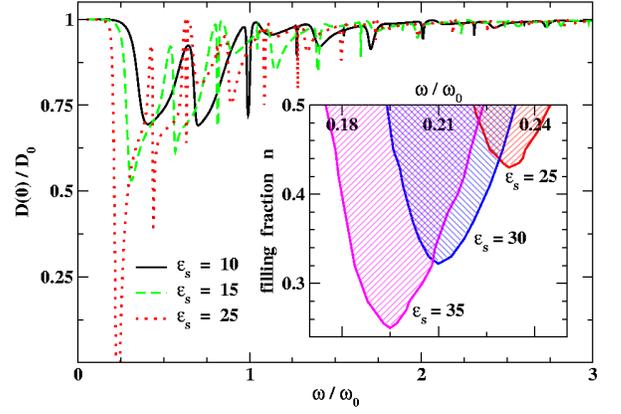} 
\end{center} 
\vspace*{-0.4cm}
\caption{(Color online) 
Normalized diffusion constant $D(0)$ for 
$\epsilon_b=1$ and scatterer volume fraction $\nu=50\%$. 
The frequency unit in all figures is 
$\omega_{\mbox{\tiny 0}} = 2\pi c/r_{\mbox{\tiny 0}}$  
with $r_{\mbox{\tiny 0}}$ the scatterer radius.
The localization transition is reached for $\epsilon_s\gtrsim 24$.
Inset: phase diagram
in the $(\nu,\omega)$ plane within the first Mie resonance
for $\epsilon_b=1$ and various $\epsilon_s$. Shaded areas:
localized phases. 
}
\label{plot1} 
\end{figure} 
We have
carefully analyzed $\gamma^{\omega}_{{\vec{q}}^{\,\prime}{{\vec{q}}}}$ 
for the selfconsistent calculation of 
$M(\Omega )$, \cite{Vol80,Kro90} exploiting time reversal symmetry of 
propagation in the active medium.
In the long-time limit ($\Omega\to 0$) the dominant contributions to 
$\gamma^{\omega}_{{\vec{q}}^{\,\prime}{{\vec{q}}}}$ are the same 
maximally crossed diagrams (Cooperons) as for conserving media,  
however now acquiring the absorption (gain)-induced decay (growth) rate 
$\xi_a^{-2}D$. $M(\Omega)$ reads   
\begin{eqnarray} 
M(\Omega ) &=& 
- \frac{(2v_{{\mbox{\tiny E}}}c_p)^2\ 
u_{\epsilon} \left[ 
2\pi \omega u_{\epsilon} N(\omega ) + r_{\epsilon}A_{\epsilon}  
- 2\omega^2\I\epsilon_b 
\right] 
}
{\pi \omega N(\omega) D_0 D(\Omega)} 
\nonumber\\ 
&& 
\hspace*{-1.5cm} 
\times   
\left\lmoustache \!\frac {{\rm d}^3 q}{(2\pi)^3} \right. 
\left\lmoustache \!\frac {{\rm d}^3 q'}{(2\pi)^3}\right. 
\frac{ 
[\vec{q}\cdot\hat{Q}] 
|\I G_q| \left(\I G_{q'}\right)^2 
[\vec{q}\,'\cdot\hat{Q}] 
} 
{\frac{-i\Omega}{D(\Omega)} + \left(\vec{q}+\vec{q}\,' \right)^2 +  
\xi_a^{-2} }\, . 
\label{MD} 
\end{eqnarray} 
Eqs.\ (\ref{D_omega_full})-(\ref{MD})  
constitute the selfconsistency equations for the diffusion coefficient 
$D(\Omega )$ and the correlation length $\xi _a$ in presence of absorption or 
gain. We have evaluated them below employing the independent scatterer  
approximation of Ref.\ \onlinecite{Tig93} for the single-particle quantities.  
More sophisticated approximation schemes, 
suitable for high scatterer concentrations,  
like the coherent potential approximation (CPA), \cite{Bus95} will 
be considered elsewhere. 

\section{Discussion}

Let us first discuss the kernel $M(\Omega )$. It describes the enhanced  
backscattering. For a conserving medium  
($\I\epsilon_b=\I\epsilon_s=0$), where $\xi _a^{-2}=0$, it drives the AL 
transition due to its negative, infrared divergent contribution.  
\cite{Vol80,Kro93} For illustration we show in Fig.\ \ref{plot1}  
the AL phase diagram  and $D(\Omega=0)$ for $\epsilon_b=1$ and various real 
values of $\epsilon_s$, displaying strong suppression near the Mie resonances,
albeit for rather high dielectric contrast $\Delta\epsilon$.   
\begin{figure}[t] 
\begin{center} 
\includegraphics[width=0.9\linewidth]{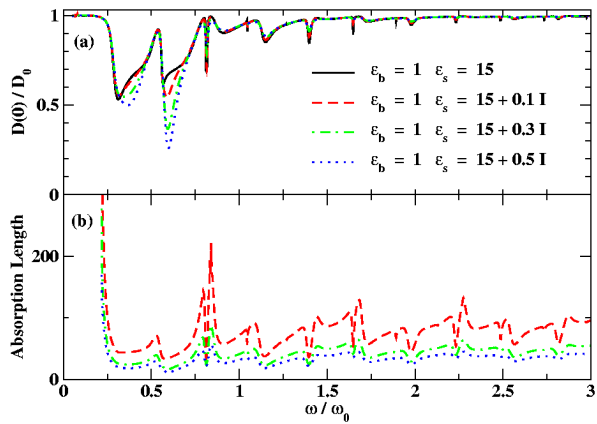} 
\end{center} 
\vspace*{-0.4cm}
\caption{(Color online) 
(a) Diffusion constant $D(0)$ for  $\nu =50\%$. 
The scatterers 
exhibit different  amounts of absorption as indicated. 
(b) Absorption length $\xi_a$  
in units of the scatterer radius 
for the systems in (a). 
} 
\label{plot2} 
\end{figure} 
Introducing now absorption or gain, the quadratic inverse   
correlation length $\xi_a^{-2}$ becomes non-zero and can 
assume both positive and negative values. True AL due to repeated 
enhanced backscattering is no longer possible in this case, because, 
through selfconsistency, a vanishing $D(0)$ would drive  
$\xi_a^{-2}\to\infty$ and $D(0)M(0)\to 0$.  
Inspection of $M(\Omega)$ shows analytically that to leading order
in $\xi_a^{-2}$ the real part of  
the integral in Eq.\ (\ref{MD}) is symmetrical with respect to  
$\xi_a^{-2} \leftrightarrow -\xi_a^{-2}$,  
i.e. AL is suppressed by absorption or gain in a  
symmetrical way. This is in agreement with the surprising  
absorption/gain duality of Ref.\  
\onlinecite{Beenakker96}, which is valid for small gain and short times.  
 
However, we do find important deviations from this result 
for systems where there is no  
symmetry between background and scattering medium (like in our $d=3$ 
case):  
(1) As seen from Eq.\ (\ref{xi_a}), $\xi_a^{-2}$ itself is not  
symmetric in the sign of $\I \epsilon _b$ or $\I \epsilon _s$ and  
can be positive even for purely emissive media,  
$\I \epsilon _b<0$, $\I \epsilon _s<0$. 
(2) The full diffusion coefficient has additional contributions $D_b$, $D_s$
from loss/gain in the background and in the scatterers, Eqs.\ (\ref{D_B}), 
which to our knowledge have not been reported before.  
While $D_b$ is always positive for absorption and negative for gain as 
expected, $D_s$ has a complicated dependence on the signs of  
$\I \epsilon _b$, $\I \epsilon _s$ and depends sensitively on whether  
absorption and/or gain occurs in the background medium or in the scatterers.  
This is because, in contrast to 
$D_b$, the impurity scattering contribution $D_s$ results from an intricate  
interplay between elastic momentum relaxation and absorption/gain processes.  
We emphasize that the existence and the form of $D_b$, $D_s$, and $\xi_a^{-2}$ 
are a direct consequence of the non-conserving terms in the WI 
Eq.\ (\ref{Ward_2}), and, thus, are exact. 

\begin{figure}[t]
\begin{center}
\includegraphics[width=0.9\linewidth]{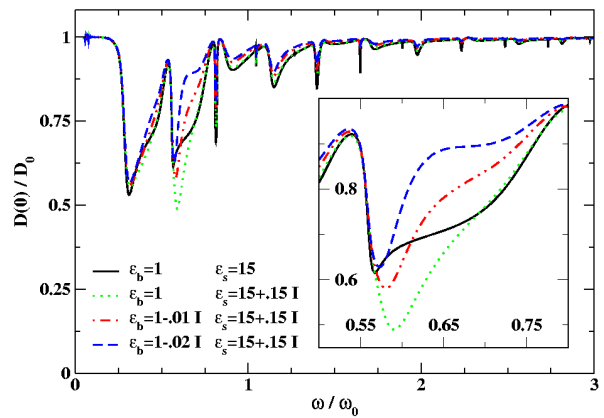}
\end{center}
\vspace*{-0.4cm}
\caption{(Color online)
Diffusion constant $D(0)$ for absorbing scatterers ($\nu=50\%$) 
in a host medium with optical gain, as indicated.
Parameters are chosen such that $\xi_a^2>0$.
}
\label{plot3}
\end{figure}

The complete scenario of localization effects is now as follows. 
Even though in the presence of absorption or gain there are no true 
Anderson localized modes because of the finite $\xi_a^{-2}$,  
the contributions $D_b$ and $D_s$ can strongly suppress the total 
diffusion coefficient $D(0)$. This is shown in Fig.\ \ref{plot2} 
for a system of absorbing scatterers embedded in air.  
Even moderate absorption drastically 
decreases the diffusion constant $D(0)$ close to the low-order Mie resonances.
At the same frequencies the correlation length $\xi_a$ is suppressed even more  
dramatically. For the case of a purely absorbing system, $\xi_a$ may be 
identified with the effective absorption length for diffusive modes.  
For the case of emission, e.g. in the background, two scenarios are possible. 
(i) $\xi_a^{-2}>0$; this may occur due to the subtle  
interplay with momentum relaxation processes and absorption in the 
scatterers. Both $D(0)$ and the intensity correlation length $\xi_a$ are 
real and finite, 
but suppressed near the Mie resonances, as shown in Fig.\ \ref{plot3}. 
It is also seen that, e.g., absorption in the scatterers 
can be partially compensated by emission in the background. 
(ii) \R\, $\xi_a^{-2}<0$; this is realized for sufficiently strong gain. 
It implies a pole on the real axis in the integration range 
of Eq.\ (\ref{MD}) and, hence, complex $D(0)$ and $\xi_a$.  
Fourier transforming Eq.\ (\ref{P_E}), this means an 
exponential intensity growth for long times with rate 
$1/\tau_a=\R [(\xi_a^{-2}+Q^2)D(0)]$, 
in qualitative agreement with Ref.\ \onlinecite{Soukoulis99}, 
thus reconciling these long-time results with the weak gain or short-time
results of Ref. \ \onlinecite{Beenakker96}. This growth is
modulated by temporal and spatial oscillations with characteristic
frequency $\Omega_D= - Q^2\I D(0)$ and wave number  $k_D=\I 1/\xi_a$, 
respectively.
We interpret this oscillatory behavior as a memory effect, 
originating from the competition between the
enhanced backscattering of waves and their amplified propagation in the 
surrounding medium. Selfinduced oscillations have been found before in
other driven systems with competing dynamics.

\section{Conclusion}

In conclusion, we have presented a semianalytical theory for the  
interplay of strong localization effects and absorption or  
gain. True AL is not
possible in the presence of either loss or gain. However,
strong renormalizations of the diffusion constant $D$ arise
from the violation of the conservation laws.
These renormalizations depend sensitively on whether 
absorption/gain occurs in the scatterers or the background.
Intimately connected with the suppression of $D$ is the appearance and
reduction of a finite intensity correlation length $\xi_a$, even though
there are no truly localized modes. 
$\xi_a$ includes effects of both, impurity scattering and loss/gain in the 
medium, and, thus, can be shorter than the scattering mean free path
$\ell$. For example, in a pure medium with loss/gain $\xi _a$ 
would characterize the absorption/gain length and would be finite, while 
$\ell$ would obviously be infinite. 
It should be emphasized that the present theory incorporates both,
the physics of multiple impurity scattering valid at length scales 
larger than $\ell$ (diffuson and Cooperon contributions)
and the physics of coherent amplification 
present at all length scales. Therefore, it is expected to 
give an accurate estimate for $\xi_a$. 
We conjecture that in random lasers this finite length scale $\xi_a$ 
might define the coherence volume necessary for 
resonant feedback, that is observed experimentally \cite{Cao01STAT}
and that appears to be smaller than $\ell$ in those experiments.
In order to substantiate this conjecture, the present localization
theory should be coupled selfconsistently to the laser rate equations,
which will yield results for the position dependent dielelectric
function above the lasing threshold ($\I \epsilon _{b,s} \neq 0$)
and, hence, for the lasing mode volume.  
This will be subject of further research. Oscillatory behavior
in space and time is also predicted for sufficiently strong gain,
although it is presumably difficult to observe, as it occurs only 
during the exponential intensity growth between the laser threshold
and saturation. 
 
Useful discussions with 
M. Arnold, H. Cao, D. Chigrin, B. Shapiro, and C.M. Soukoulis are  
gratefully acknowledged. This work was partly supported by DFG through 
FG 557 and grant no. KR1726/3 (J.K., A.L.).

\end{document}